\documentclass[sn-basic]{sn-jnl}

\usepackage{sidecap,makecell}
\usepackage{soul}
\usepackage{hyperref}
\usepackage[mathlines]{lineno} 
\sidecaptionvpos{figure}{c}

\newcommand{\tr}{^{\sf T}}
\begin{document}

\title[One Citation, One Vote!]{One citation, one vote! \endgraf A new approach for analysing \endgraf check-all-that-apply (CATA) data \endgraf
using L1-norm methods}

\author*[1]{\fnm{C.} \sur{Chaya} }\email{carolina.chaya@upm.es}
\author[2]{\fnm{J.C.} \sur{Castura} }\email{jcastura@compusense.com}
\author[3]{\fnm{M.J.} \sur{Greenacre}}\email{michael.greenacre@upf.edu}

\affil*[1]{\orgdiv{Department of Agricultural Economics, Statistics and Business Management, Business Innovation and Management Group}, \orgname{Universidad Politécnica de Madrid}, \orgaddress{\country{Spain}}} 
\affil[2]{\orgname{Compusense Inc.}, \orgaddress{\city{Guelph}, \country{Canada}}}
\affil[3]{\orgdiv{Department of Economics and Business}, \orgname{Universitat Pompeu Fabra, and Barcelona School of Management}, \orgaddress{\city{Barcelona}, \country{Spain}}}



\abstract{
A unified framework is provided for analysing check-all-that-apply (CATA) product data following the ``one citation, one vote" principle.
CATA data arise from studies where \textit{A} assessors evaluate \textit{P} products by describing samples by checking all of the \textit{T} terms that apply.
Giving every citation the same weight, regardless of the assessor, product, or term, leads to analyses based on the L1 norm where the median absolute deviation is the measure of dispersion.
Five permutation tests are proposed to answer the following questions.
Do any products differ? For which terms do products differ? Within each of the terms, which products differ? Which product pairs differ? On which terms does each product pair differ?
Additionally, we show how products and terms can be clustered following the ``one citation, one vote" principle and how principal component analysis using the L1-norm (L1-PCA) can be applied to visualize CATA results in few dimensions.
Together, the permutation tests, clustering methods, and L1-PCA provide a unified approach that provides robust results measured in citation percentages. The proposed methods are illustrated using a data set in which 100 consumers evaluated 11 products using 34 CATA terms.
R code is provided to perform the analyses.}

\keywords{check-all-that-apply (CATA), pick-any, permutation tests, median absolute deviation (MAD), L1 norm, L1-norm principal component analysis (L1-PCA)}

\maketitle

\section{Introduction} \label{Introduction}
Sensory and consumer studies are often designed to investigate how humans perceive foods, beverages, or non-food products. These studies usually involve structured sample evaluations of the products by human assessors with data analysis by sensometric methods. 
Some studies obtain precise sensory characterizations of the products from panelists trained to identify and scale sensory attributes from a consensus lexicon. Other studies investigate how consumers perceive and conceptualise products. 
Both types of studies frequently use the check-all-that-apply (CATA) question format \citep{ares2023check, meyners2014check}, which provides a list of \textit{terms} to allow \textit{assessors} to characterise \textit{products}.
For example, a CATA question might instruct the assessor to check as many sensory or emotion terms from the list as describe the product sample under evaluation.
Such a study leads to a three-way array (assessors-by-products-by-terms) of raw data. 

Since checked terms, called \textit{citations} or \textit{elicitations}, can be analysed as count data, CATA data are often  aggregated over the assessors to obtain a two-way products-by-terms table of frequencies. 
Thus, a CATA frequency table counts the assessors who cited each term for each product.

CATA frequency tables are often analysed as contingency tables \citep{Dooley:10, mahieu2021multiple, meyners2014check, Meyners:13, valentin2012quick}, in which a normalisation or adjustment is applied that in some sense equalises the contributions of variables with different citation rates. 
The widely used normalisation inherent in the chi-squared test has this feature: each cell frequency is expressed relative to the reciprocal of the square root of its expectation, given the column and row marginal sums. 
However, a CATA frequency table is not a contingency table.
In a contingency table, each member of the assessor sample contributes to only one table cell. 
In a CATA frequency table, 
each cell 
indicates how many assessors cited a particular term (column) for a particular product (row).
For this reason, we propose to analyse raw counts without normalising terms or products. 
We will investigate differences between products for each term separately (univariately) and over all terms simultaneously (multivariately), using the novel approach of treating each citation equally. 
Our proposal is embodied in the following principle: one citation, one vote! 

This proposal provides a new way of summarising and analysing CATA data that we have not seen used previously. 
Treating every citation equally leads us to use medians (not means) as our measure of central tendency, medians of absolute deviations from medians as our measure of dispersion (not variances), and median absolute deviations (not Euclidean distances) to quantify differences when grouping products and grouping terms in cluster analysis. Additionally, we use L1-norm principal component analysis (L1-PCA; \citealp{Ke:05}), rather than conventional L2-norm PCA (L2-PCA), to summarise the product information visually in few dimensions.

As we will show, treating all citations equally gives many advantages.  
Terms are not rescaled, as in some other analyses, so infrequently cited terms having small product differences are not adjusted to carry the same weight as frequently cited terms having large differences between products. 
These analyses yield results that are easy to interpret and communicate. 
A drawback of our approach is we cannot test hypotheses in the classical manner (e.g., chi-squared tests), but we overcome this disadvantage using resampling procedures (e.g. permutation tests, bootstrapping). 

Data and methods are introduced in Section \ref{MaterialMethods}. 
The data come from a study involving 100 consumers who used 34 emotion terms to evaluate 11 blackcurrant squash products (fruity beverages) available commercially in the United Kingdom \citep{Ng:13a}. 
There, we also describe the proposed statistical methods, all of which are aligned with the ``one citation, one vote" principle. We describe how to conduct distribution-free permutation tests, as well as how to adjust for multiple testing. 
We describe how to perform cluster analysis of products and terms, as well as how to explore relationships between products and terms using L1-PCA. 
Section \ref{Results} presents the results obtained from applying these methods to the blackcurrant squash CATA data. We discuss these results and make conclusions in Section \ref{Discussion}.

\section{Materials and Methods} \label{MaterialMethods}
\subsection{Data from the blackcurrant squash study} \label{squashes}
This paper uses data extracted from research by \cite{Ng:13a} in which consumers evaluated 11 commercial blackcurrant squashes, representing a range of sensory and packaging properties observed in the UK market. 
The original labelling of the 11 squash products (\textsf{P1} to \textsf{P11}) in \cite{Ng:13a} is supplemented here by two additional letters (\textsf{ab}). 
Letter \textsf{a} indicates the economical, niche and standard market-value categories, respectively \textsf{a} = \textsf{e, n, s}.
Letter \textsf{b} indicates products with and without added sugar, respectively  \textsf{b} = \textsf{w, o}.
For example, the first product \textsf{P1sw} is a standard product with added sugar.

A preliminary study described by \cite{Ng:13b} was conducted to select relevant conceptual terms with emotional or functional connotations \citep{Thomson2010linking} for evaluating blackcurrant squashes under three conditions: blind, pack, and informed. 
In this manuscript, we analyse results from the blind condition only, in which consumers evaluated each sample without seeing the package.
In this condition, consumers' emotional conceptualizations of the samples are not influenced by information about brand, assumptions about the market-value category (niche vs standard vs economical), added sugar, or use of non-nutritive sweeteners, if any.

Results from the preliminary study were used to finalise a CATA questionnaire. Assessors were tasked with checking all terms that described how they felt immediately after tasting each sample.
The 34 emotion terms of the blind condition can be found in Table \ref{BJ1}.

In total, $A=100$ assessors evaluated the $P=11$ products according to the $T=34$ terms.
Products were presented by experimental design to balance sample order effects. 
Terms were presented by experimental design to balance term position effects.

\begin{SCfigure}[][b]
\includegraphics[width=5.5cm]{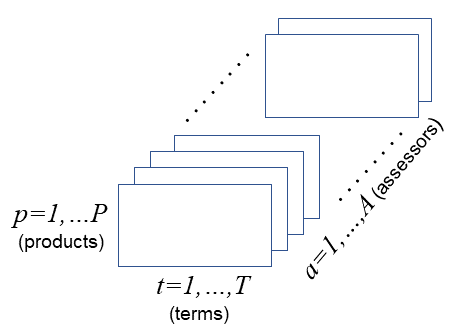}
\caption{Structure of raw CATA data in the form of a three-way array, where cell $(p,t,a)$ has the value {1} if assessor $a$ cited term $t$ for product $p$; otherwise, the cell value is {0}. The three-way array is 
shown as a series of $A$ tables  corresponding to the assessors. The summation of these $P\times T$ tables over all the $A$ assessor slices yields the $P\times T$ CATA frequency table of counts. For CATA percentages, the counts in each cell are expressed as a percentage of $A$.}
\label{DataStructure}
\end{SCfigure}

The three-way raw data, depicted schematically in Fig. \ref{DataStructure} contains indicator variables: zeros indicate terms not cited; ones indicate terms cited.
From these raw data, the CATA frequency table is the $P \times T$ table formed by summing the $A$ slices across the assessors.

\subsection{A new approach, simple and robust}
\label{NewApproach}

As described in Section \ref{Introduction}, when considering how to analyse CATA data, we began with the following premise: each citation has exactly the same importance, regardless of the term elicited, the product evaluated, or the assessor who made the judgment. 
Since every assessor evaluates all products using the same terms, the frequency scale (i.e., counts) and the percentage scale (i.e., expressing each cell as a percentage of $A$) are equivalent. 
For ease of reference, the table of citation percentages will be called a CATA table.

Many previous approaches for analysing CATA data are based on classical statistics and involve squared differences, which are used to compute variances, least-squares estimates, chi-squared statistics, and so on. 
Even a simple statistic such as the arithmetic mean, which estimates an expected value, is based on least-squares estimation: the mean is the value that minimises the sum of squared differences to the sample values. The mean is sensitive to outliers because least-squares estimation gives greater weight to more-distant sample values and lesser weight to more-central ones.
For example, the counts 10 and 12 have the absolute difference 2 and the squared difference 4, whereas the counts 10 and 14 have the absolute difference 4 and the squared difference 16. 
Squaring amplifies larger differences more than smaller differences, which is inconsistent with the ``one citation, one vote" principle. 

To treat every response equally, we will measure differences by their absolute values and use the median as a measure of centrality.
The median is the value that minimises the sum of absolute differences to the sample values \citep{Schwertman:90}. 
We measure dispersion about the median by finding the median of the absolute deviations between the sample values and the median,
called the \textit{median absolute deviation} (MAD) about the median.
Since absolute differences are less sensitive to outliers, the median and MAD are more robust measures of central tendency and overall difference, and fit the ``one citation, one vote" principle.

If ${\bf E}$ is a CATA (percentage) table, 
then each element $e_{pt}$ gives the percentage of the $A$ assessors who cited term $t$ for product $p$. In this case, 
$\textsf{M}(t) \equiv \textsf{M} \{ e_{pt} \, ; \, p=1,\ldots,P \}$ denotes the median citation percentage for term $t$ across the $P$ products, i.e., the median of column $t$ of matrix {\bf E}. 
The MAD for term $t$ is defined as the median of absolute deviations from the median:
\begin{equation}
{\rm MAD}(t) \equiv \textsf{M}\{ \vert e_{pt}-\textsf{M}(t) \vert \, ; \, p=1,\ldots,P \}
  \label{MAD}
\end{equation}
For example, in the CATA table that we will analyse soon, the following percentage values are observed for the set of 11 products for the two terms {\it Happy} and {\it Sickly}:
\smallskip

\begin{tabular}{lrrrrrrrrrrr}
  {\em Happy} (\%) & 54 & 31 & 15 & 22 & 40 & 25 & 55 & 28 & 41 & 15 & 48\\
  {\em Sickly} (\%) & 21 & 25 & 19 & 15 & 12 & 30 & 12 & 35 & 17 & 14 & 12
\end{tabular}

\smallskip
\noindent
\textit{Happy} has the median percentage 31\% across the products, while \textit{Sickly} has the median percentage 17\%.  
Within each set, we calculate the MAD dispersion values using the sample values and their respective medians. The dispersions are  
10\% for the term \textit{Happy} and 5\% for the term \textit{Sickly}. 

We draw attention to three points. 
First, as just illustrated, the measures of central tendency and dispersion usually increase in tandem. When least-squares methods are applied in multivariate analyses, terms having unequal variances are often rescaled to equalise their contributions to the analysis; for example, in correspondence analysis \citep{Benzecri:73, Greenacre:16a}, the chi-square normalisation rescales variables and samples so all are on the same footing.
In our case, following the ``one citation, one vote" principle, 
we do not attempt to balance the influence of any terms, products, or assessors. 
Larger differences between products based on citations of the term \textit{Happy} are considered to discriminate the products better than the smaller differences between products based on citations of the term \textit{Sickly}. 
Second, the summary values used (median, MAD) are very easy to understand and interpret because they are measured in percentages, which is exactly the same scale used in the CATA table.
Third, the median and the MAD
are more robust to outliers than measures based on squared differences, such as the mean and variance, which are influenced more strongly by outlying sample values.

When it comes to hypothesis testing, instead of using classical statistics with known, usually asymptotic theoretical distributions based on squared differences and least-squares methods, we will take a distribution-free approach, described next.

\subsection{Permutation testing} \label{Permutation}

To test differences between products on the selected terms, we make use of the simple, non-parametric approach called permutation testing; for a complete treatment, see \cite{Good:04}, and for previous applications in the analysis of CATA tables, see 
\cite{Meyners:13} and \cite{mahieu2021multiple}. As we will show, this approach can be used to test various hypotheses, such as whether individual cells in the CATA table are larger or smaller than expected or whether a pair of products differs on any of the terms. The approach naturally generalises to multivariate tests of whether products differ globally, accounting for all terms simultaneously.

The null hypothesis assumes that there is no difference between products. If the null hypothesis is true, then any product differences have occurred by chance. We can simulate a potential outcome under this null hypothesis by permuting product labels within each assessor, leading to a new CATA table. 
The permutation procedure requires the raw data and is repeated many times, each time producing another random CATA table. 
In this study, we have generated $B=9\,999$ random CATA tables, which generate a null distribution of the chosen test statistic.
These CATA tables represent a range of potential outcomes that could have occurred by chance assuming the products in fact elicit the same set of terms.
The chosen test statistic is computed for the observed CATA table and for each of the $B$ simulated CATA tables, giving a null distribution comprising $B+1$ values in total.
Suppose the value of the test statistic from the observed table is so extreme that we almost never see values as or more extreme. 
The low probability of observing such an extreme outcome, indicated by a very small p-value, leads us to conclude that the null hypothesis is improbable.
Based on this result, we would conclude that products differ with a high level of confidence, as indicated by the small p-value estimated from the null distribution. 

Now, we describe the procedure in more specific detail. Operationally, each simulation takes the form of random permutations of the product labels within each assessor. 
Fig. \ref{DataStructure} shows the set of citations in the $P\times T$ matrix ``slice" of the array corresponding to a particular assessor.
Permuting the $P$ rows of this matrix assigns rows to different product labels but preserves dependencies in elicitations across the multivariate terms. For example, a particular assessor who tends to endorse the terms \textit{Pleased} and \textit{At~ease} together or not at all has data dependencies that are kept intact by permuting the rows (products) keeping the columns (terms) intact.
This procedure is applied to data from all $A$ assessors. Since rows are assigned to product labels independently per assessor, any two assessors will usually not have the same product rows assigned to the same product labels. 
After permuting product rows within assessors, assessor matrix slices are aggregated into a simulated $P\times T$ CATA table.
Repeating this procedure $B$ times yields $B$ random CATA tables, which together with the original CATA table, gives $B+1$ CATA tables in total. 

Suppose we want to evaluate the evidence against a null hypothesis that products have the same citation percentage for a particular term $t$. 
The test statistic chosen to quantify the observed product differences is the MAD statistic given by Eq. (\ref{MAD}),
that is, the median absolute deviation from the $P$ products to their median value.
The null distribution for term $t$ is comprised of the MAD statistic from the observed CATA table and $B$ MAD values from the random CATA tables. 
The p-value is the proportion of these $B+1$ MAD values 
that are greater than or equal to the observed MAD statistic. 
The MAD statistic and corresponding p-value can be obtained for each of the $T$ terms in the same manner. 

Since the observed MAD statistic is part of the null distribution, at least one MAD value from the null distribution always meets this criterion. For this reason, if a null distribution contains $B+1=10\,000$ MAD values, the smallest possible p-value ($0.0001$) is obtained if the observed MAD statistic is larger than all other MAD values in the null distribution. 

The beauty of permutation testing is that the researcher can select any suitable test statistic, even one with an unknown statistical distribution. 
Since we have generated $B=9\,999$ random CATA tables, a p-value can be estimated to four decimal places.
Performing more permutations would give even more precise p-values.
In the present context, we propose the permutation tests described in Table \ref{tbl_PermTests}. The testing procedure is straightforward and, in many cases, permutation results obtained for one test can be re-used in another test.

\begin{table*}[h]
{\small
\begin{tabular}{p{0.08\linewidth} | p{0.15\linewidth} | p{0.35\linewidth} | p{0.15\linewidth}}
Test \#  & Type & Question addressed & Max.~tests \\
\hline
\centering\arraybackslash 1 & Multivariate & Do any product citation percentages differ across the terms? & \centering\arraybackslash 1 \\
\centering\arraybackslash 2 & Univariate & Do any product citation percentages differ in term $t$? & \centering\arraybackslash $T$ \\
\centering\arraybackslash 3 & Elementwise & Does the product $p$ citation percentage differ from the median citation percentage in term $t$? & \centering\arraybackslash $PT$ \\
\centering\arraybackslash 4 & Pairwise\newline multivariate & Does this pair of products differ in citation percentages across the terms? & \centering\arraybackslash $\frac{1}{2}P(P-1)$ \\
\centering\arraybackslash 5 & Pairwise\newline
univariate & Does this pair of products differ in citation percentages of term $t$? & \centering\arraybackslash $\frac{1}{2}P(P-1)T$ \\
\hline
\end{tabular}
}
\smallskip
\caption{Permutation tests proposed. The column `Max.~tests' gives the maximum number of tests for a data set with $P$ products and $T$ terms.}
\label{tbl_PermTests}
\end{table*}

Tests in Table \ref{tbl_PermTests} are interrelated. The MAD statistic for each elementwise test (Test 3) is the absolute difference of the percentage of product $p$ on term $t$ to the median percentage of term $t$. The MAD statistic for each univariate test (Test 2) is the median of the $P$ elementwise MAD statistics for term $t$ in Test 3. The MAD statistic for the multivariate test (Test 1) is the median of the $T$ univariate MAD statistics in Test 2.
Additionally, the MAD statistic for each univariate paired product comparison (Test 5) is the absolute difference between the two products on term $t$. The MAD statistic for the multivariate paired comparison of these two products (Test 4) is the median of the $T$ MAD statistics from these univariate paired comparisons of these two products in Test 5.

The multivariate test (Test 1) evaluates whether there are global product differences, considering all terms. A significant test result encourages further investigations.
When making these additional investigations, multiple hypotheses are tested, introducing statistical considerations that will be discussed in the next section. For this reason, we will discuss the issue of multiple testing in Section \ref{MultipleTesting}, then further describe Tests 2--5.


\subsection{Multiple testing and false discovery}
\label{MultipleTesting}
In a null hypothesis test, a true null hypothesis is rejected with Type I error rate $\alpha$ set by the researcher. If multiple independent hypotheses are each tested at level $\alpha$, the probability of making one or more Type I errors can be higher, or even substantially higher, than $\alpha$. 
For this reason, many researchers adjust their protocol when conducting multiple tests.
In this section, we discuss how multiple testing adjustments will be applied in the analysis of the CATA data.

Suppose we investigate each term to determine if citation percentages differ significantly from the median on each of the $T=34$ terms (Test 2). 
If the products are, in fact, the same with respect to all terms, the probability of making at least one Type I error is $1-(1-0.05)^{34}=0.825$, far larger than the nominal rate $0.05$.
Had there been $100$ terms, we would make at least one Type I error with probability $0.994$---a near certainty.

Since controlling the Type I error rate across all tests reduces statistical power severely, researchers often elect to control the false discovery rate (FDR) instead.
A rejected null hypothesis, called a ``discovery", can be either true or false. 
The discovery is true (i.e., correct) if the rejected null hypothesis is, in fact, false. 
The discovery is false (i.e., incorrect) if the rejected null hypothesis is, in fact, true. 
When FDR=$0.05$, the proportion of false discoveries among all discoveries is controlled at level $0.05$.

We are not conducting independent null hypothesis tests of independent predictor variables; rather, our tests are dependent hypothesis tests of correlated response variables.
For this reason, it could be argued there is no reason to control the FDR.
In this case, our rationale for controlling the FDR is simply to restrict communication of results to terms that merit the most attention.

Controlling the FDR for $M$ tests using the Benjamini-Hochberg (BH) step-up procedure \citep{Benjamini:95} involves sorting the p-values in ascending order alongside the arithmetic series of $M$ BH values, $\{\,\frac{1}{M}0.05,\ \frac{2}{M}0.05, \ldots, \ \frac{M-1}{M}0.05,\  0.05\,\}$ = $\left(\{\,1,2,\ldots,M\,\}/M\right) \cdot 0.05$, starting at $\frac{1}{M} 0.05$ (corresponding to the conservative Bonferroni test), and increasing by increments of $\frac{1}{M} 0.05$ to $0.05$ (corresponding to the Type I error rate). 
The BH critical value is the largest p-value that is smaller than its corresponding BH value. 
All p-values less than or equal to this critical value are declared significant. 
These results can be visualised in a heatmap colour-coded to emphasise the significant p-values. 
This BH procedure can be applied to Tests 2--5 in Table \ref{tbl_PermTests}.
In each case, we use FDR=0.05, but each BH series differs according to the maximum number of hypotheses tested, given in column `Max.~Tests' of Table \ref{tbl_PermTests}.


Apart from Tests 1--3 on the CATA table values themselves, it is also of interest to test their pairwise differences, univariately or multivariately.
The multivariate product paired comparison Test 4 evaluates which of the $\frac{1}{2}P(P-1)$ product pairs differ significantly across the terms.
The test statistic for a particular product pair is the median of the $T$ absolute differences in citation percentages between the two products in the CATA table. 
Using the permutation procedure, we obtain the corresponding MAD values between the two products computed on the $B$ random CATA tables. 
These $B$ simulated MAD values, together with the test statistic, constitute the null distribution. 

As before, the p-values are obtained from the null distribution and  evaluated using the BH procedure, then visualised in a products-by-products ($P\times P$) heatmap.

Doing the pairwise test for each term $t$ (Test 5),
the null distribution is comprised of the test statistic, which is the observed signed difference, along with the similar differences in the $B$ random CATA tables.
The p-value is again the proportion of absolute outcomes in the null distribution as large or larger than the absolute test statistic.
It is possible to test only product pairs determined to differ significantly based on the multivariate test (Test 4).
Here, all product pairs on all terms are tested using the BH procedure, and the results are visualised for each term in a $P \times P$ heatmap.

\subsection{Hierarchical cluster analysis}
\label{L1HCA}  
A researcher might want to cluster products into groups where products in the same group are conceptualised similarly and products in different groups are conceptualised differently. 
We perform a hierarchical clustering of the products using complete-linkage clustering, which tends to yield relatively tight, widely separated clusters since the distance between the furthest member of each group is considered whenever clusters are merged.
For consistency with the ``one citation, one vote" principle, we measure the dissimilarity of any pair of products using the median of the pairwise MAD values from the $T$ terms, which is the test statistic from Test 4 in Table \ref{tbl_PermTests}. 

A researcher might also want to cluster terms into groups where terms in the same group are cited in a similar manner and terms in different groups are cited differently. 
To do so, we centre the CATA table by subtracting the column medians, then apply complete-linkage clusters based on the paired comparisons of the terms.
The distance between any pair of terms is the median of the pairwise MAD values from the $P$ products in this centred table. 


\subsection{L1-norm principal component analysis (L1-PCA)}
\label{L1PCA}  
Without loss of generality, suppose matrix $\bf X$ has median-centred columns. 
L1-PCA decomposes $\bf X$ into two matrices: a score matrix $\bf T$ and loading matrix $\bf P$. Dimension reduction is possible since retaining only $K$ components minimises the sum of absolute residuals
\begin{equation}    
\|{\bf X}-{\bf T}_{K}{\bf P}_{K}\tr\|_1
\label{sumResiduals}
\end{equation}
where $\tr$ denotes matrix transposition and $\|\cdot\|_1$ denotes the L1 norm, that is, the sum of absolute elements.
Several L1-PCA algorithms have been proposed for approximating ${\bf X}$ with ${\bf T}_{K}{\bf P}_{K}\tr$. We use the algorithm proposed by \cite{Ke:05}.

Additionally, we show the uncertainty of the products in the L1-PCA visualization.
To do so, we conduct bootstrapping at the assessor level.
Each assessor has $PT$ values in his or her set of responses (a slice of the array shown in Fig. \ref{DataStructure}.) 
Assessors are randomly chosen with replacement, until another three-way array of $A$ slices, as shown in Fig. \ref{DataStructure}, is constructed, where some assessors are included more than once, some not at all. 
From this bootstrapped array another CATA table is formed, and its rows are projected onto the original solution already established using linear programming to minimize the sum of absolute deviances, subject to constraints \citep{Ke:05}.
The L1-PCA algorithm we used \citep{Ke:05} is robust to initialization values and reaches more precise solutions by allowing more iterations towards a smaller convergence tolerance. A solution could be considered converged if changing the initialization values, allowing more iterations, and specifying a smaller convergence tolerance does not provide meaningful improvement.
After repeating this bootstrap procedure 1000 times, each time projecting the rows to the solution, the points are overlaid with 95\% covering ellipses, which estimate the confidence regions for the products.
If confidence ellipses on two product means do not overlap, this is an indication, but not conclusive proof, of a significant difference between the products.
In the present L1 norm framework, we prefer to judge this pairwise product difference using the permutation test previously proposed. 

The total dispersion is the L1 norm of $\bf X$. The L1-PCA with $K$ components extracts the largest possible sum of absolute deviances from $\bf X$. The proportion of the L1 norm extracted in these $K$ components, 
\begin{equation}
    \textsf{prop}_K = \frac{\|{\bf T}_{K}{\bf P}_{K}\tr\|_1}{\|{\bf X}\|_1} 
    \label{ExplainedProportion}
\end{equation}
quantifies the success of this L1-PCA in approximating $\bf X$.
The solutions in L1-PCA are not nested as in regular PCA, so percentages of explanation cannot be ascribed to individual dimensions, only to specific $K$-dimensional solutions.
To obtain a scree plot similar to that usually produced in PCA, the L1-PCA is performed separately for solutions of dimension $K=1,2,\ldots,{\rm min}\{P,T\}$, and the gains in explanation are then computed, going from the solutions of dimension $K-1$ to $K$ (where, for $K=0$, the explanation is 0).
Different researchers will make different choices for how many components to retain. Since we were interested in product separation, we kept only the first $K$ components that separated products, i.e., where one or more product confidence ellipse excluded the origin.

\subsection{Software}\label{Software}
Data analyses were conducted in R 4.2.2 \citep{R:2024}. 
Functions \texttt{madperm()} and \texttt{mad.dist()} in the R package \textbf{cata} \citep{castura_cata} were used to conduct permutation tests and to calculate MAD-based distances for cluster analyses. 
R code in \hypertarget{SupplMaterialLink1}{\hyperlink{SupplMaterialLabel}{Supplementary~Material}} shows how to reproduce the permutation tests presented here. 
Running these permutation tests took approximately about four minutes on a laptop running 64-bit Windows 10 Pro 22H2 with an Intel\textsuperscript{\textregistered} Core\texttrademark{} i5-1235U, 1300 Mhz, 10 Core(s), 12 Logical Processor(s) and 16 GB RAM.
Functions \texttt{l1pca()} and \texttt{l1projection()} in the R package \textbf{pcaL1} \citep{Jot:2023} were used to conduct L1-PCA and to project bootstrapped sample points onto the L1-PCA biplots.
Function \texttt{CIplot\_biv()} in the R package \textbf{easyCODA} \citep{Greenacre:18} used these points to calculate and draw confidence ellipses.
Heatmaps and dendrograms were drawn using the function \texttt{pheatmap()} in the R package \textbf{pheatmap} \citep{kolde:2019}.

\section{Results} \label{Results}

\subsection{Tests and heatmaps} \label{Subsection_TestsHeatmaps}
In the blackcurrant squash data set, counts and percentages are identical since $A=100$ consumers evaluated the products. 

\begin{SCfigure}[][h]
\includegraphics[width=6.5cm]{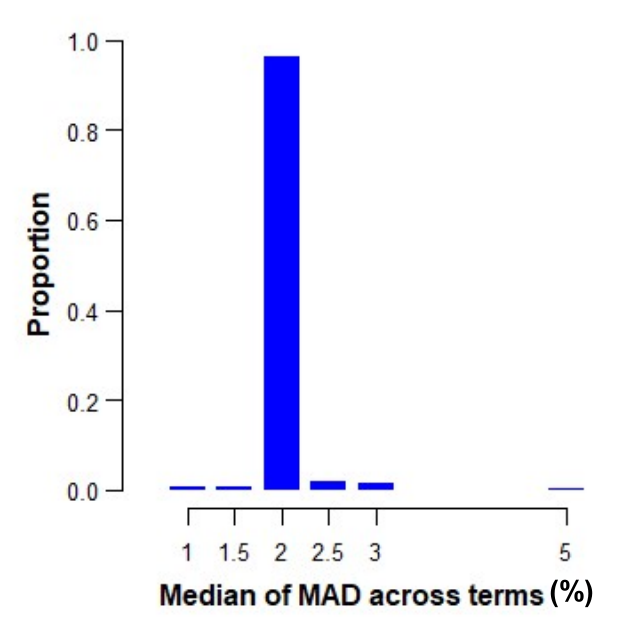}
\caption{ 
Null distribution values for the overall test statistic for product differences (x axis) vs proportion of occurrence (y axis). The median absolute deviance (MAD) is calculated for each term; the observed test statistic is the median of the MAD values across the 34 terms, which was 5\% (at the extreme right).} 
\label{CATAoverall}
\end{SCfigure}

Test 1 conducts a global multivariate test to determine whether product differences exist on any term. The test statistic is the median of the MAD values from the $34$ terms, which was 5\% (at the extreme right)..
Fig. \ref{CATAoverall} shows the null distribution for this test statistic.


A high proportion (0.96) of the null distribution has the median value (2.0\%). There are relatively few values at 1.0\%, 1.5\%, 2.5\% and 3.0\%. 
The value of 5\%, which is the observed test statistic, is alone at the extreme right. 
Since none of the randomised values equal or exceed the observed value, the global multivariate test indicates significant global differences ($p=0.0001$) and justifies further analyses.

\begin{figure*}[b]
\begin{center}
\includegraphics[width=11.8cm]{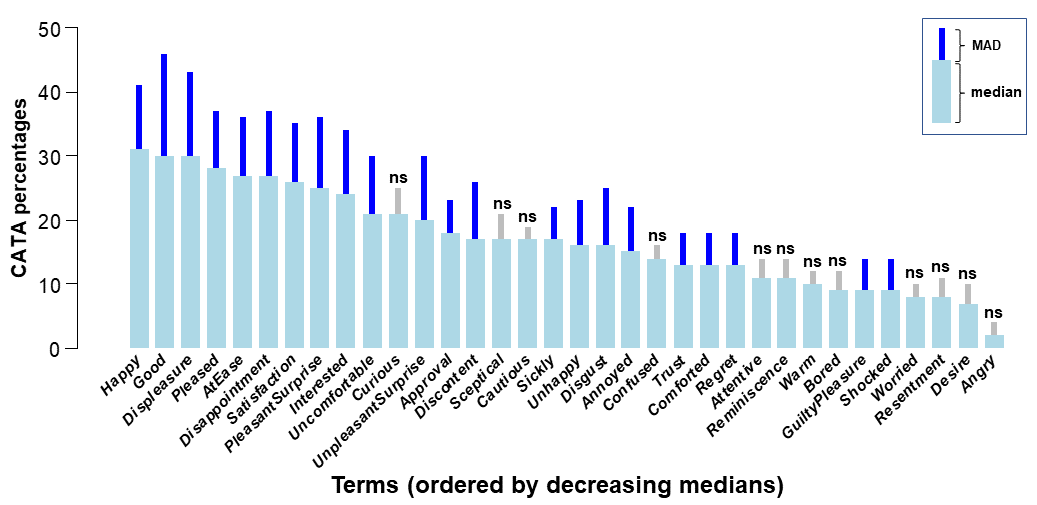}
\caption{For each term, the median citation percentage of the 11 products is shown (wide vertical bars) with the median absolute deviation (MAD) of each product represented by a thin vertical bar on top. The MAD line is shown in dark blue for the 22 significant terms and in light grey (denoted `ns') for the 12 non-significant terms.} 
\label{CATAeffects}
\end{center}
\end{figure*}

Next, we investigate which of the $T$ terms discriminate the products. 
For the term \textit{Happy}, citation percentages range from 15\% to 55\%.
The MAD value for \textit{Happy} is 10\%, which is the median ``distance" of the 11 citation percentages from the median of 31\%, on the percentage scale.
Fig. \ref{CATAeffects} shows graphically both the median (as a wide vertical bar) and MAD (as a narrow bar on top) for each term.
This plot shows the median and MAD values are correlated.
We emphasise the MAD bars of terms that significantly discriminate the products, as determined by the univariate test results (Test 2), which are described next.


To evaluate univariate test results, we conducted Test 2 permutation tests (Section \ref{Permutation}), from which we obtained 34 p-values.
These results are presented in Table \ref{BJ1}, alongside the BH series used to control the FDR at 0.05 (Section \ref{MultipleTesting}). 
The 15 terms achieving the smallest possible p-value ($p=0.0001$) are presented in no particular order. 
The critical value ($0.0125$) was obtained from term \textit{Sickly}, which had the largest p-value that was smaller than its BH step-up value. 
This critical value 
controls the FDR at level 0.05. 
The 22 terms on which the product differed significantly had MAD values of 5\% and larger, whereas the 12 other terms had MAD values of 4\% and smaller (Table \ref{BJ1}).

\begin{center}
\begin{table*}[t]
{\small
\begin{tabular}{lccc}
\textsf{Term} &  \textsf{MAD (\%)} & \textsf{Permutation p-value} & \textsf{BH step-up value}\\
\hline
\textit{Happy} & 10 & 0.0001 & 0.0015 \\
\textit{Unhappy} & 7 & 0.0001 & 0.0029 \\
\textit{Uncomfortable} & 9 & 0.0001 & 0.0044 \\
\textit{At~ease} & 9 & 0.0001 & 0.0059 \\
\textit{Pleasant~surprise} & 11 & 0.0001 & 0.0074 \\
\textit{Unpleasant~surprise} & 10 & 0.0001 & 0.0088 \\
\textit{Disappointment} & 10 & 0.0001 & 0.0103 \\
\textit{Satisfaction} & 9 & 0.0001 & 0.0118 \\
\textit{Discontent} & 9 & 0.0001 & 0.0132 \\
\textit{Disgust} & 9 & 0.0001 & 0.0147 \\
\textit{Interested} & 10 & 0.0001 & 0.0162 \\
\textit{Good} & 16 & 0.0001 & 0.0176 \\
\textit{Displeasure} & 13 & 0.0001 & 0.0191 \\
\textit{Annoyed} & 7 & 0.0001 & 0.0206 \\
\textit{Pleased} & 9 & 0.0001 & 0.0221 \\
\textit{Shocked} & 5 & 0.0002 & 0.0235 \\
\textit{Guilty~pleasure} & 5 & 0.0003 & 0.0250 \\
\textit{Regret} & 5 & 0.0011 & 0.0265 \\
\textit{Trust} & 5 & 0.0013 & 0.0279 \\
\textit{Comforted} & 5 & 0.0060 & 0.0294 \\
\textit{Approval} & 5 & 0.0094 & 0.0309 \\
\textit{Sickly} & 5 & 0.0125 & 0.0324 \\
\hline
\hline
\textit{Resentment} & 3 & 0.0704 & 0.0338 \\
\textit{Sceptical} & 4 & 0.0719 & 0.0353 \\
\textit{Desire} & 3 & 0.0793 & 0.0368 \\
\textit{Curious} & 4 & 0.1059 & 0.0382 \\
\textit{Attentive} & 3 & 0.1121 & 0.0397 \\
\textit{Angry} & 2 & 0.1573 & 0.0412 \\
\textit{Reminiscence} & 3 & 0.1891 & 0.0426 \\
\textit{Bored} & 3 & 0.1982 & 0.0441 \\
\textit{Worried} & 2 & 0.4821 & 0.0456 \\
\textit{Warm} & 2 & 0.6587 & 0.0471 \\
\textit{Confused} & 2 & 0.7350 & 0.0485 \\
\textit{Cautious} & 2 & 0.7961 & 0.0500 \\
\hline
\end{tabular} }
\smallskip
\caption{Results for the univariate test (Test 2) controlling the false discovery rate at 0.05 show sorted p-values alongside the Benjamini-Hochberg (BH) series. The 22 terms having a p-value below the critical value (0.0125) were considered significant. The non-significant terms below the double line did not discriminate the products.}
\label{BJ1}
\end{table*}
\end{center}

\vspace{-0.4cm}

\begin{figure*}[b]
\begin{center}
\vspace{-0.5cm}
\includegraphics[width=11.8cm]{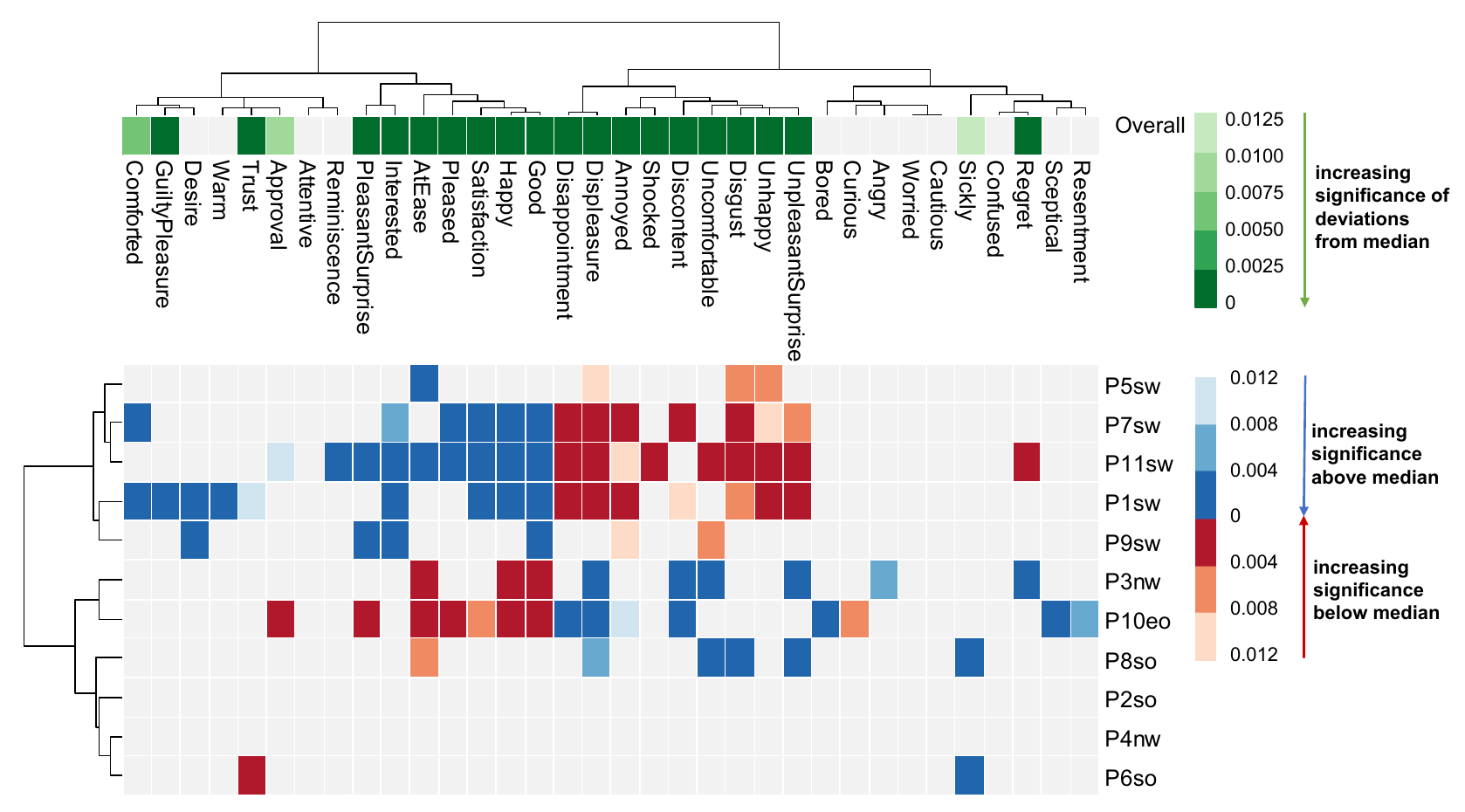}
\caption{Heatmap of p-values from univariate tests (Test 2 results shown in top row interpreted with uppermost legend) and elementwise tests (Test 3 results shown in lower matrix interpreted with the two lower legends ). Complete-linkage cluster analysis results are shown for clustering of terms (top) and clustering of products (left). In Test 2 and in Test 3, the false discovery rate was 0.05. [Products are uniquely numbered (\textsf{P1} to \textsf{P11}). The first letter indicates the market-value category: economical (e), niche (n), standard (s). The second letter indicates with (w) or without (o) added sugar.]} 
\label{HeatMapCells}
\end{center}
\end{figure*}

Fig. \ref{HeatMapCells} (top row) shows a heatmap visualisation of the univariate (Test 2) p-values in Table \ref{BJ1}. 
Terms that significantly discriminate the products are shown in green, with deeper colours indicating a more significant test result. 
Terms above the BH critical value, which are non-significant across the products, are shown in light grey.



Cluster analysis of the terms, as described in Section \ref{L1HCA}, organises the terms into a two-cluster solution, shown at the top of Fig. \ref{HeatMapCells}. Positively valenced terms (cluster 1) are contrasted against negatively valenced terms (cluster 2).
Term cluster 1 consists of 15 positively valenced terms, 11 of which discriminated the products: \textit{Comforted}, \textit{Guilty~pleasure}, \textit{Trust}, \textit{Approval}, \textit{Pleasant~surprise}, \textit{Interested},  \textit{At~ease}, \textit{Pleased}, \textit{Satisfaction}, \textit{Happy} and \textit{Good}. The lexicon developers did not classify the term \textit{Guilty~pleasure} as either positively or negatively valenced \citep{Ng:13a}, but these term clusters suggest that consumers conceptualised Guilty pleasure as positively valenced. The remaining four terms in this cluster (\textit{Desire}, \textit{Warm}, \textit{Attentive}, \textit{Reminiscence}) did not discriminate the products.
Term cluster 2 is comprised of 19 negatively valenced terms, of which 11 discriminated the products significantly (\textit{Disappointment}, \textit{Displeasure}, \textit{Annoyed}, \textit{Shocked}, \textit{Discontent}, \textit{Uncomfortable}, \textit{Disgust}, \textit{Unhappy}, \textit{Unpleasant~surprise}, \textit{Sickly}, \textit{Regret}) and eight did not (\textit{Bored}, \textit{Curious}, \textit{Angry}, \textit{Worried}, {Cautious}, \textit{Confused}, \textit{Sceptical}, \textit{Resentment}).

Fig. \ref{HeatMapCells} (lower part) also shows a $P\times T$ heatmap visualization of the elementwise (Test 3) p-values obtained by analysing, for each term, which of the $P$ products differs from the median value. 
We decided a priori to conduct all $P\times T = 11\times 34 = 374$ elementwise tests (Section \ref{Permutation}) with the FDR controlled at level 0.05 (Section \ref{MultipleTesting}).
In 374 elementwise tests, 89 p-values were less than or equal to the BH critical value of  $0.0118$.

Significant elementwise p-values were colour-shaded, where the cell colour was determined by the sign of the difference between the observed and expected count: a product significantly below the term median was coloured red; a product significantly above the term median was coloured blue. Products that were not significantly different from the median are coloured light grey.

At least one product differed significantly from the median for all but four terms: \textit{Attentive}, \textit{Worried}, \textit{Cautious}, and \textit{Confused}.
These {four terms are among} the 12 non-significant terms (coloured light grey in the top row of Fig. \ref{HeatMapCells}).
Inspection of the columns for the other eight non-significant terms show at least one significant elementwise test.
This result shows that individual products can differ significantly from the median of a term, even if the product citation percentages do not differ significantly overall on that term.

Fig. \ref{HeatMapCells} shows the dendrogram from the complete-linkage cluster analysis of products to the left of the elementwise heatmap. The rows (products) are sorted to accommodate this dendrogram which shows a crisp two-cluster solution. 
Product cluster 1 is composed of five standard products with added sugar (\textsf{P5sw}, \textsf{P7sw}, \textsf{P11sw}, \textsf{P1sw}, \textsf{P9sw}). Product cluster 2 is composed of six more varied products: four products without added sugar (\textsf{P10eo}, \textsf{P8so}, \textsf{P2so} and \textsf{P6so}) and two niche products with added sugar (\textsf{P3nw} and \textsf{P4nw}).

The elementwise heatmap reveals approximate characterisations of products in the two product clusters. 
Members of product cluster 1 elicited significantly higher citation percentages than the median product in positively valenced terms in term cluster 1 and significantly lower citation percentages than the median product in negatively valenced terms in term cluster 2.
By contrast, members of product cluster 2 elicited significantly higher citation percentages than the median product in negatively valenced terms in term cluster 2 and significantly lower citation percentages than the median product in positively valenced terms in term cluster 1. 

We draw attention to selected product results (rows) in the elementwise heatmap related to product cluster 2. 
Like other members of product cluster 2, the economical product \textsf{P10eo} had higher-than-median citation percentages for negatively valenced terms in term cluster 2 and lower-than-median citation percentages for positively valenced terms in term cluster 1. 
Additionally, \textsf{P10eo} elicited significantly higher-than-median citation percentages for \textit{Bored}, \textit{Sceptical}, and \textit{Resentment} and a significantly lower-than-median citation percentage for \textit{Curious}.
Unlike the other 10 products, \textsf{P10eo} had relatively high intensities of catty, confectionery blackcurrant and artificial sweetness \citep{Ng:12}, which might explain why this product elicited these negatively valenced terms.

Three standard products without added sugar (\textsf{P2so}, \textsf{P6so} and \textsf{P8so}) were also allocated to cluster 2. 
Besides of the general trend of product cluster 2, \textsf{P6so} and \textsf{P8so} elicited higher-than-median citation percentages for \textit{Sickly} in term cluster 2. 
In a previous study, \textsf{P6so} and \textsf{P8so} were observed to have the highest intensities of artificial sweetness and natural processed blackcurrant \citep{Ng:12}, suggesting the relatively high \textit{Sickly} citation rates were due to cloying oversweetness and processed flavour. 
These same sensory properties could be the reason for a lower-than-median citation percentage of \textit{Trust} for \textsf{P6so}.  

Compared to other products, \textsf{P3nw} was more intense in green/leafy and earthy flavours, more bitter, and more astringent \citep{Ng:12}. 
If consumers in the blind sensory evaluation did not expect these sensory properties, it could explain why this niche product elicited negatively valenced terms more often than other products. 
Individual tests (Tests 3 and 5) can be significant even when an overall test (Test 2) is not. Some researchers might enforce greater consistency by conducting Test 3 only for terms that were significant based on Test 2, but conducting all tests helps in the interpretation of the consumers' response elicited by \textsf{P3nw}, adding some cues for the identification of specific differences.
Although products did not differ significantly in citation percentages for \textit{Angry} based on the univariate test results (Test 2; Table \ref{BJ1}, Fig. \ref{CATAeffects}), \textsf{P3nw} elicited higher-than-median citation percentages for \textit{Angry} based on elementwise tests (Test 3; Fig. \ref{HeatMapCells}) and univariate pairwise tests (Test 5; results not shown).
Again, these results might be explained by the sensory profile of \textsf{P3nw} mentioned above.

Test 4 investigates which product pairs differ considering all $T$ terms simultaneously. 
For each of the $55$ paired comparisons, the test statistic is the median absolute paired difference across the $T$ terms.
Permutation testing (Section \ref{Permutation}) yields a p-value for each product pair.
The critical value that controlled FDR at 0.05 (Section \ref{MultipleTesting}) was $0.0281$.
Based on this critical value, 34 paired comparisons were significantly different. 
Table \ref{Pairwise} shows the multivariate paired comparison test statistics, emphasising these significant differences. 
The rows and columns of this table are ordered as in Fig. \ref{HeatMapCells}.

\setlength{\tabcolsep}{4pt}
\begin{center}
\begin{table*}[t]
{\small
\begin{tabular}{lrrrrrrrrrrr}
 & \textsf{P5sw} & \textsf{P7sw} & \textsf{P11sw} & \textsf{P1sw} & \textsf{P9sw} & \textsf{P3nw} & \textsf{P10eo} & \textsf{P8so} & \textsf{P2so} &  \textsf{P4nw} \\
 \hline
\textsf{P7sw}   & 3.0 \\                                            
\textsf{P11sw}  & 3.0 & 2.0 \\                                        
\textsf{P1sw}   & 5.0 & 2.5 & 4.0 \\                                    
\textsf{P9sw}   & 4.0 & 4.0 &  4.0 & 4.0 \\                               
\textsf{P3nw}  & {\bf 12.5} & {\bf 14.5} & {\bf 14.5} & {\bf 15.5} & {\bf 12.0} \\                         
\textsf{P10eo} & {\bf 11.0} & {\bf 14.0} & {\bf 14.5} & {\bf 16.5} & {\bf 13.0} & 4.0 \\                   
\textsf{P8so}  & {\bf 10.0} & {\bf 12.5} & {\bf 12.5} & {\bf 12.0} & {\bf 8.5} & 5.0 & {\bf 5.5} \\              

\textsf{P2so}  & {\bf 8.0} & {\bf 10.0} & {\bf 10.5} & {\bf 10.0} & {\bf 7.0} & {\bf 6.0}  & {\bf 8.0} & 4.5 \\ 

\textsf{P4nw}  & {\bf 8.5} & {\bf 10.5} & {\bf 12.5} & {\bf 11.0} & {\bf 9.0} & 4.0 & \bf{5.5} & 4.5 & 4.0 \\   

\textsf{P6so}  & {\bf 10.0} & {\bf 13.0} & {\bf 14.0} & {\bf 13.5} & {\bf 10.5} & 4.0 & 4.5 & 4.0 & 4.0 &  2.0 \\
\hline
\end{tabular}
}
\smallskip
\caption{Median absolute difference in citation percentages across all terms from each of the 55 multivariate product paired comparisons. The false discovery rate was 0.05. The 34 significantly different pairs are shown in boldface.}
\label{Pairwise}
\end{table*}
\end{center}

\vspace{-0.4cm}

\begin{SCfigure}[][t]
\includegraphics[width=8cm]{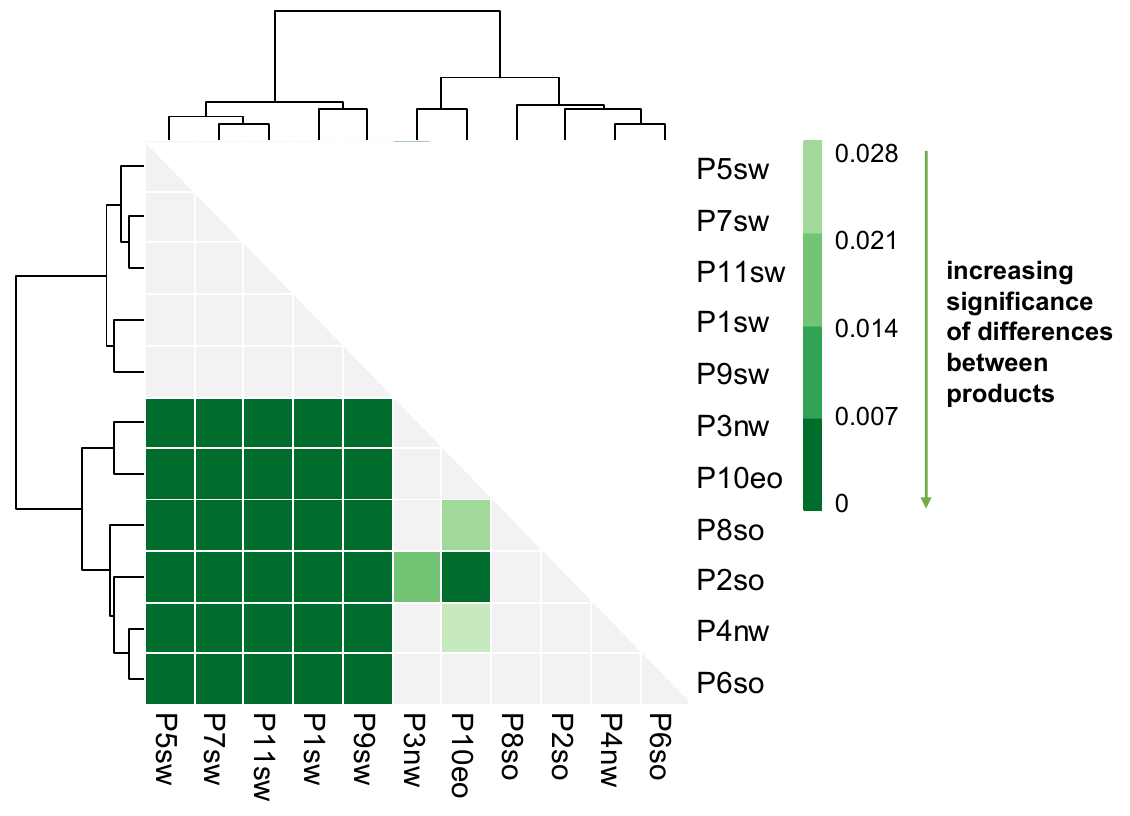}
\caption{Heatmap of p-values for the 55 multivariate product paired comparisons in Table \ref{Pairwise} (Test 4). With the false discovery rate controlled at 0.05, 34 of the 55 pairs were significantly different (shown in green). The diagonal and cells for non-significant tests results are shown in light grey.}
\label{HeatMapDiffs}
\end{SCfigure}

Fig. \ref{HeatMapDiffs} shows the dendrogram from the complete-linkage cluster analysis of products on horizontal and vertical axes of this heatmap, the same dendrogram that appeared in Fig. \ref{HeatMapCells} on the left side.
The heatmap combined with the dendrogram facilitates the interpretation. 
It shows standard products with added sugar (product cluster 1) differ significantly from products without added sugar and niche products (product cluster 2). There was no evidence of significant paired differences within product cluster 1, but various paired differences were found within product cluster 2. 
Among products without added sugar, the economical product \textsf{P10eo} had a different emotional profile than the standard products \textsf{P8so} and \textsf{P2so}. The two niche products, \textsf{P3nw} and \textsf{P4nw}, elicited a different emotional response than products \textsf{P2so} and \textsf{P10eo}, respectively, from the same product cluster 2.

Test results also showed \textsf{P10eo} differed significantly from one niche product (\textsf{P4nw}) and two standard products without added sugar (\textsf{P2so} and \textsf{P8so}). 
Although we can venture reasonable guesses why each of these product pairs differed by referring to the elementwise test results (Fig. \ref{HeatMapCells}), a better understanding is obtained from univariate product paired comparisons, which will be evaluated next.

Each univariate paired comparison test (Test 5) investigates whether a term discriminates the product pair.
Since this study is exploratory but also illustrative, we decided a priori to investigate all product pairs across all terms.
Each test statistic is the signed difference in citation percentages between the two products.
Using the permutation test procedure (Section \ref{Permutation}), we obtained $TP(P-1)/2=1870$ p-values, which were placed alongside the BH arithmetic series to obtain the BH critical value $0.0130$. 
Using this critical value to control the FDR at level 0.05, we obtained 487 significant test results based on Test 5.

To illustrate the univariate paired comparison test, we return to \textit{Happy} and \textit{Sickly}, two terms introduced in Section \ref{NewApproach}.
Both these terms significantly discriminated the products (Table \ref{BJ1}). 
As expected, \textit{Happy} ($p=0.0001$) discriminated the products better than \textit{Sickly} ($p=0.0125$), and discriminated more product pairs based on the Test 5. 
We found 24 product pairs differ significantly in \textit{Happy} citation percentages and 13 product pairs differ significantly in \textit{Sickly} citation percentages.

\begin{figure*}[b]
\begin{center}
\includegraphics[width=11.8cm]{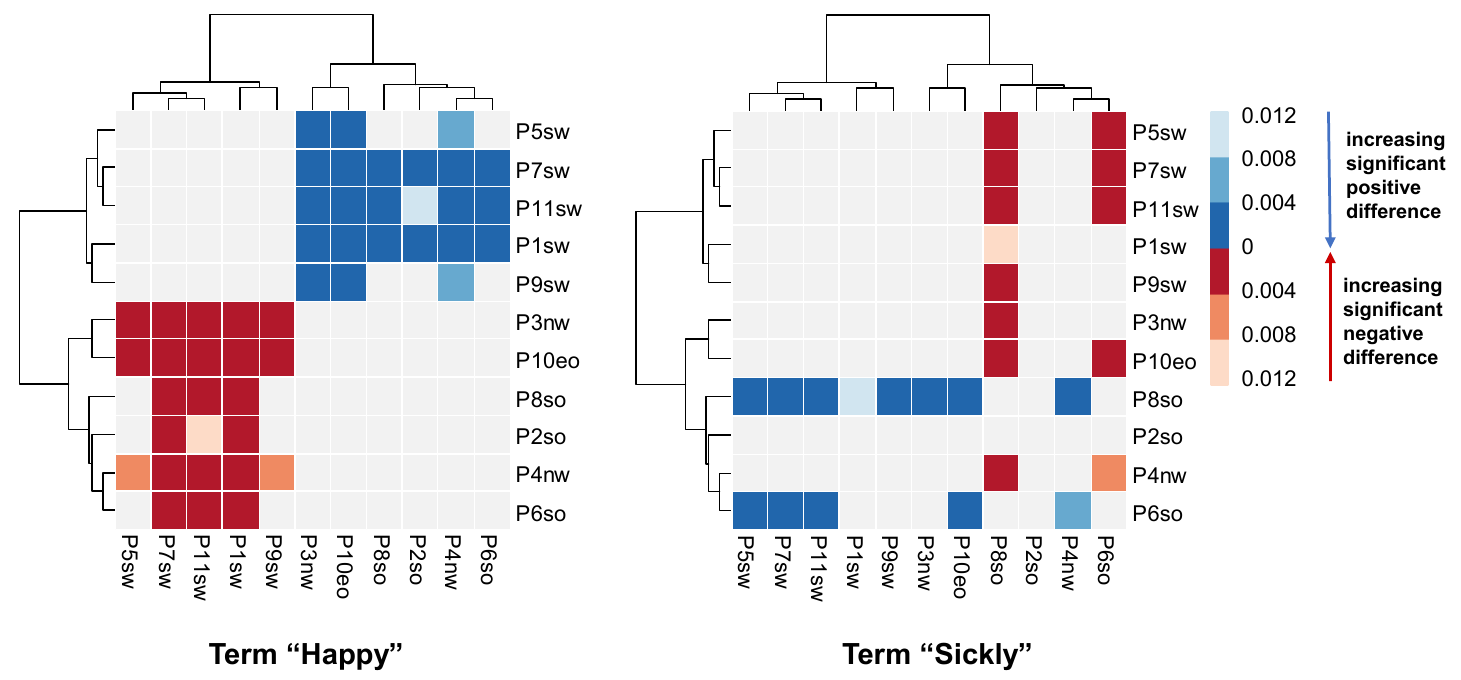}
\caption{Results for univariate paired comparisons of products on \textit{Happy} and \textit{Sickly} with the false discovery rate controlled at 0.05 across all such tests. The diagonal and cells for non-significant tests results are shown in light grey. Blue and red cells indicate significant differences; see legend for details.}
\label{HappySickly}
\end{center}
\end{figure*}

The p-values for \textit{Happy} and \textit{Sickly} are visualised in heatmaps in Fig. \ref{HappySickly}.
In these heatmaps, the direction of the difference in citation percentages is obtained by subtracting the column product from the row product (i.e. rows minus columns), where positive and negative values are colour-coded blue and red, respectively, for product pairs that differ significantly. 
The differences above and below each main diagonal have the opposite sign but the same absolute magnitude, hence the blue--red symmetry across the diagonal.
Products are ordered and shown together with the dendrograms as in Fig. \ref{HeatMapDiffs}, which show the results from the complete-linkage cluster analysis of products.

Results show nearly all members of product cluster 1 have significantly higher \textit{Happy} citation percentages than members of product cluster 2. 
No products in cluster 2 had a significantly higher citation rate than any product in product cluster 1, but some other within-cluster differences were observed.
The observed \textit{Happy} citation percentages were larger in product cluster 1 than in product cluster 2. Within product cluster 1, the observed \textit{Happy} citation percentages were highest for \textsf{P7sw} and lowest for \textsf{P5sw}, but these differences were not significant. Due to these differences, we detected more (6) significant paired differences between \textsf{P7sw} and the products in product cluster 2 than detected (3) between \textsf{P5sw} and products in product cluster 2. 
\textsf{P3nw} and \textsf{P10eo} had the lowest \textit{Happy} citation percentages (both 15\%); both these products differed significantly from all five products in product cluster 1.

Paired comparison results for \textit{Sickly} are driven entirely by two standard products without sugar from product cluster 2. 
\textsf{P6so} and \textsf{P8so} had \textit{Sickly} citation percentages of 35\% and 30\%, respectively, whereas citation percentages for other products lie between 12\% and 25\%.
\textsf{P6so} had a significantly higher citation percentage than five other products, three in product cluster 1 and two in product cluster 2. 
\textsf{P8so} had a significantly higher citation percentage than eight other products, including all members of product cluster 1 and three products in product cluster 2.

\subsection{Dimension reduction using L1-PCA}
\begin{figure*}[t]
\begin{center}
\includegraphics[width=10.5cm]{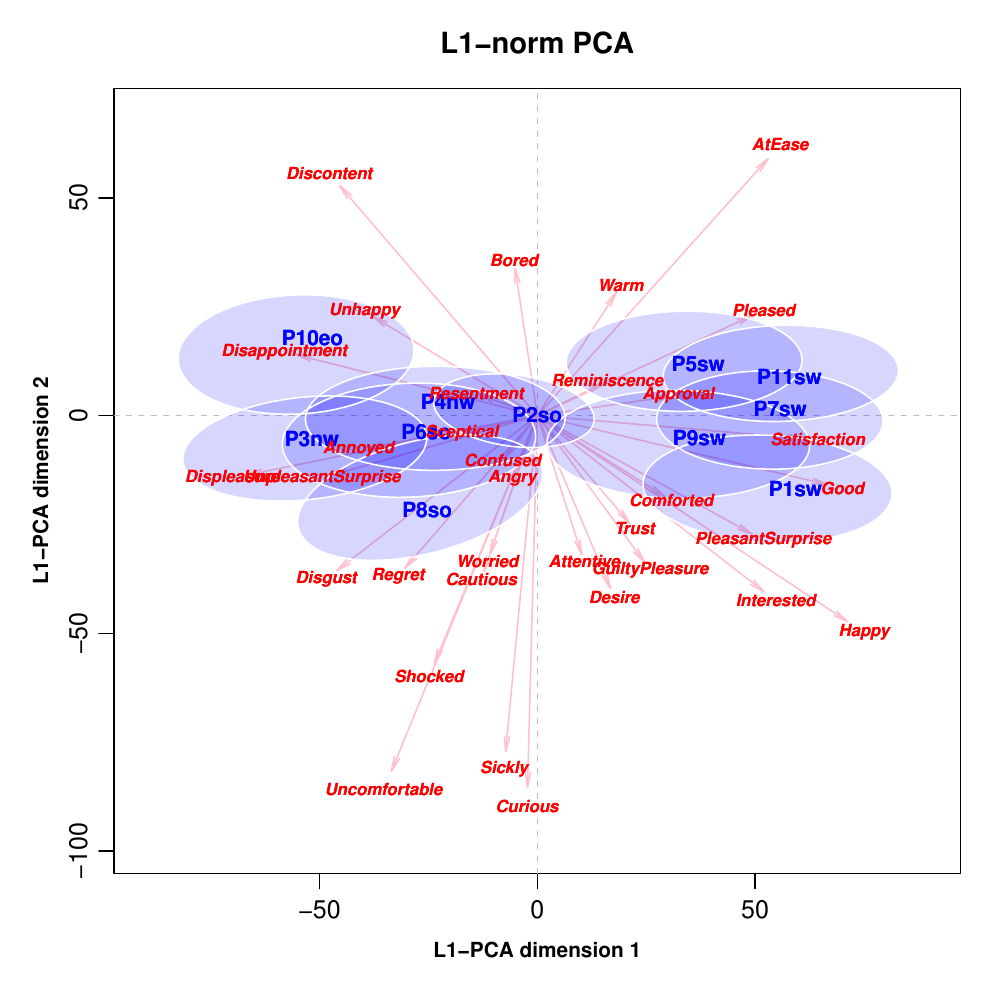}
\caption{L1-PCA biplot of the CATA table, with 95\% confidence ellipses for the product medians.}
\label{L1pca}
\end{center}
\end{figure*}

The L1-PCA biplot (Fig. \ref{L1pca}) gives a compact summary of the data structure and shows several results that have been observed in the permutation tests.
The scale of these two components is percentage units. 
The biplot shows scores (products, as points) visualised together with the loading vectors, which are rescaled for legibility.
The origin coincides with product \textsf{P2so}, which lies at the median.

\begin{SCfigure}
\includegraphics[width=8.2cm]{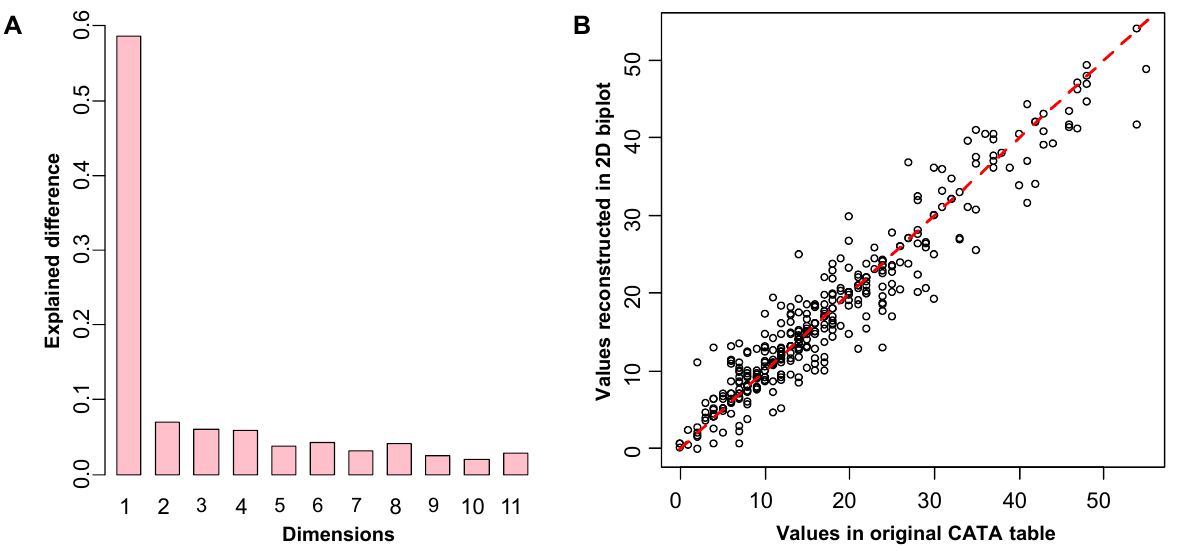}
\caption{A. Scree plot of the L1-PCA, showing the gains in explaining the CATA table for solutions of increasing dimension ($K$). B. Scatterplot of CATA table percentages (horizontal axis) and the values reconstructed from the two-dimensional L1-PCA biplot of Fig. \ref{L1pca}.}
\label{ReconstructionPlot}
\end{SCfigure}
\noindent
Based on Eq.~(\ref{ExplainedProportion}), the proportion of the CATA table explained in two components is 0.657.
The proportion of the CATA table explained by one L1-PCA component is 0.586, so the gain in the two-dimensional solution is 0.071.
Fig. \ref{ReconstructionPlot}A shows the equivalent of a scree plot for the L1-PCA, that is, the gains in explanation of the CATA table for increasing dimensions.
To show the quality of the L1-PCA two-dimensional solution, the concordance between the CATA table percentages $\bf E$ and the estimated values obtained from the two-dimensional L1-PCA solution is shown in the scatterplot of Fig. \ref{ReconstructionPlot}B. 
Notice in Eq.~(\ref{sumResiduals}) that $\bf X$ is median-centered, hence the medians have to be added to ${\bf T}_{K}{\bf P}_{K}\tr$ to obtain the PCA estimates for the solution in $K=2$ dimensions.

The five standard products with added sugar (i.e., with codes ending \textsf{-sw}) are aligned with attributes such as \textit{At~ease}, \textit{Good}, \textit{Happy}, \textit{Interested}, \textit{Pleasant~surprise}, and \textit{Satisfaction}.
The attributes \textit{Disappointment}, \textit{Displeased}, \textit{Unpleasant surprise} and \textit{Unhappy} are aligned with products \textsf{P10eo} and \textsf{P3nw}, attributes that also tend to characterise the products \textsf{P4nw} and \textsf{P6so}, but since their 95\% confidence ellipses overlap the origin, it cannot be concluded that their sensory profiles differ from the median sensory profile, at least based on the results in this plane.

In most cases, the significant multivariate paired product differences (Fig. \ref{HeatMapDiffs}) have non-overlapping ellipses in the L1-PCA plane shown (Fig. \ref{L1pca}).
For example, both Test 4 and the L1-PCA indicate particular product pairs differ:  \textsf{P10eo} and \textsf{P2so}, \textsf{P3nw} and \textsf{P2so}, and \textsf{P10eo} and \textsf{P8so}.
In Fig. \ref{L1pca}, ellipses for \textsf{P2so} and \textsf{P8so} do not overlap with the ellipse for \textsf{P10eo} (which had catty notes), even though all of these products were in product cluster 2. 

The five standard products with added sugar are well separated from the other six products, with the exception of slightly overlapping ellipses for \textsf{P9sw} and \textsf{P2so}, and \textsf{P9sw} and \textsf{P4nw}, which are better separated in higher dimensions.
Table \ref{Pairwise}, however, confirms the strict separation of these two product clusters statistically.

In some cases, interpretations from the two analyses differ. 
For example, the multivariate pairwise test (Fig. \ref{HeatMapDiffs}) did not detect a significant difference between \textsf{P1sw} and either \textsf{P5sw} or \textsf{P11sw}, even though their 95\% confidence ellipses are separated in the first L1-PCA plane (Fig. \ref{L1pca}).
The discrepancy probably occurs because the multivariate pairwise test considers all attributes simultaneously, whereas the L1-PCA results in this plane considers only the linear combinations of attributes in the two components shown. 

\section{Discussion} \label{Discussion}

\subsection{Constraining the number of tests} \label{Discuss_constrain}
In many CATA studies, all assessors evaluate each product once using the same CATA terms. A CATA frequency table from such studies has the unique property that each cell is the result of a separate counting of the same sample of respondents. 
This property suggests a wide variety of hypothesis tests. 
We have listed five such tests from which the researcher can choose. 

Researchers can constrain which tests are conducted so only relevant aspects are investigated. 
For example, three researchers might apply Test 5 in different ways depending on the study objectives. One researcher might apply Test 5 only to terms that differ significantly based on Test 2 results. A second researcher might apply Test 5 only to product pairs that differ significantly based on Test 4 results. A third researcher might apply Test 5 only to the significant terms based on Test 2 results for only the significant product pairs based on Test 4 results.


\subsection{Comparison with conventional tests} \label{Discuss_comparison}
Among the proposed permutation tests in Table \ref{tbl_PermTests}, the univariate test (Test 2) and the univariate pairwise test (Test 5) are respectively analogous to the Cochran's Q test \citep{cochran:1950, Meyners:13} and McNemar's test \citep{McNemar:49, Meyners:13}.
Results from these two tests are provided in \hypertarget{SupplMaterialLink2}{\hyperlink{SupplMaterialLabel}{Supplementary~Material}} to facilitate comparisons with permutation testing results in Section \ref{Results}.

Under a true null hypothesis that $P$ products elicit the same citations proportion, Cochran's Q test statistic follows the asymptotic chi-squared distribution with $P-1$ degrees of freedom. 
Cochran's Q test detected product differences in 30 terms, more than were detected using the permutation testing approach (22).
Although controlling FDR will tend to reduce the number of significant test results, in this data set, the number of significant Cochran's Q test results was the same with and without FDR control, and the number of significant results from Test 2 was also the same with and without FDR control (Supplementary Material, Table S1).

We also conducted McNemar's test to determine whether two products elicit the same citation proportion for each of the 55 product pairs on each of the 34 attributes.
In each of these 1870 tests, differentiating citations (i.e. in which the term is endorsed for only one of the two products) will be binomially distributed.
With FDR controlled at level 0.05, McNemar's test detected 434 significant paired differences, fewer than were reported based on the permutation testing approach (487).
When FDR was uncontrolled, McNemar's test detected 639 significant paired differences, fewer than were detected using the permutation testing approach (660).
These results show that the permutation testing approach is at least sometimes more powerful than the exact version of McNemar's test.
An advantage of McNemar's test, however, is that it returns consistent output rapidly from an exact distribution that does not require a permutation testing procedure (Supplementary Material, Table S2).

Previously, \cite{Meyners:13} also proposed a global test similar to Test 1 based on the randomisation testing procedure, which was approximated by the sum of Cochran's Q test statistics. 
This test is often highly significant in typical consumers tests, as in this data set (Fig. \ref{CATAoverall}). 

\subsection{Future studies and limitations} \label{Discuss_limitations}

The proposed permutation tests might be applied to other types of data also.
For example, quantitative sensory data collected on rating scales could be converted to percentages, then analysed using method proposed in Table \ref{tbl_PermTests}, where the L1-PCA can provide dimension reduction and summary.
Other approaches, such as taxicab correspondence analysis \citep{choulakian2006taxicab}, might also be investigated further.

The method for constructing confidence ellipses in the L1-PCA could be investigated further to determine whether these ellipses have approximately frequentist properties and whether other confidence-ellipse constructing procedures are superior \citep{castura2022discriminability, castura2023evaluation}.

Since multivariate pairwise tests (Test 4) are based on median MAD values, it seems possible that real differences could be missed if the products are strongly separated on a few terms, but a large number of non-differentiating or irrelevant terms are included in the study. 
Further study would be needed to evaluate the power of the proposed permutation tests compared to conventional tests for analysing CATA data. 

A seed should be set before conducting permutation tests to ensure reproducibility. 
Software referenced in this manuscript can facilitate such matters.
Permutation testing requires enough permutations to achieve stable results.
Our results based on 9\,999 permutations were usually close (within $\pm0.01$) to results based on 99\,999 permutations. 
Researchers are advised to determine whether results are stable by conducting additional permutations or confirming that the conclusions are the same if a different seed is used.

Neither term clusters nor product clusters seemed to be influenced by marginal citation percentages of the terms and products, respectively; however, a potential drawback with the proposed method of clustering terms is that ties might increase when there are relatively few products.
Usually, there are more terms than products, but if the number of terms is small, ties might arise when clustering products also.

\section{Conclusion} \label{Conclusion}

In this study, we tested specific hypotheses given in Table \ref{tbl_PermTests} using permutation tests that make no particular assumptions about the distribution of the data under the null distribution.
Due to the large number of tests, we opted to control the false discovery rate to restrict communication systematically to the significant test results in which we were most confident. 
This approach provided a data-driven way to determine which results to report that we found superior to simply choosing a smaller Type I error rate.

Our new framework for analysing CATA data is simple and robust.
This framework emerges from the principle that every single citation recorded in a CATA product-by-term table counts the same as any other citation, no matter the assessor, product, or term. 
This ``one citation, one vote" principle implies a different set of statistical measures based on the L1 norm, contrary to other methods based on the L2 norm.
When using the L1 norm, the measure of dispersion is based on the absolute differences in the CATA values, which are reported in percentage units, whereas methods based on the L2 norm are based on squared differences, often with some type of normalization.
In the present approach, data are analysed ``as is", meaning there are no potentially controversial data transformations.
Thus, even though there are differences in the frequencies of checking between terms and between assessors, we avoid any attempt to normalize the data, both term-wise and assessor-wise, since such normalizations would imply that citations get varying numbers of ``votes", depending on the terms and/or the assessor. 

Thanks to the absence of any transformation on the CATA table, 
the proposed analyses are often easier to understand than analyses based on the L2 norm. 
Our approach using the L1 norm introduces a considerable simplification into the analysis of CATA tables for the food researcher.

The methods described in this paper are being introduced into the R package \textbf{cata} \citep{castura_cata}, which will also include this paper's data on blackcurrant squashes.

\section*{Author contributions}
\noindent
CC: Data curation, Formal analysis, Investigation, Resources, Writing – review \& editing.

\smallskip
\noindent
JCC: Conceptualization, Formal analysis, Software, Validation, Writing – original draft, Writing – review \& editing. 

\smallskip
\noindent
MJG: Conceptualization, Formal analysis, Methodology, Software, Visualization, Writing – original draft, Writing – review \& editing.

\section*{Supplementary material}
\noindent
Supplementary material is provided in the form of two files.
\begin{enumerate}
   \item A file with five sections: (S1) R code for replicating results, (S2) Cochran's Q test results, (S3) McNemar's test results, (S4) principal component analysis and correspondence analysis, and (S5) a description of the 3D video animation of the L1-PCA results in Suppl. Video V1.
   \item Suppl. Video V1. Score plot for L1-PCA of the CATA table with three components showing 95\% confidence ellipoids for the product coordinates. The video shows rotation around the vertical second axis, so the dispersion of the product ellipsoids on the third dimension can be seen.
\end{enumerate}
\bibliography{CATA_arXiv} 

\newpage

\centerline{\textbf{\Large Supplementary Material}} 
\medskip
\centerline{\large One citation, one vote!}
\smallskip
\centerline{\large A new approach for analysing
check-all-that-apply (CATA) data}
\smallskip
\centerline{\large using L1-norm methods}
\medskip
\centerline{\large Carolina Chaya, John C. Castura and Michael J. Greenacre}

\setcounter{table}{0}
\setcounter{figure}{0}
\setcounter{page}{1}
\makeatletter
\renewcommand{\thetable}{S\arabic{table}}
\renewcommand{\thefigure}{S\arabic{figure}}

\subsection*{S1. R code for replicating results}
Permutation testing results presented in this manuscript can be replicated in R 4.4.2 \citep{R:2024} by installing the R package \textbf{cata} \citep{castura_cata}, then running the following code:\\

\noindent \texttt{require(cata)}\\

\noindent\texttt{\# permutation tests can take several minutes to finish }\\
\noindent\texttt{results <-}\\
\indent\texttt{madperm(squash, B=9999,} \\
\indent\texttt{seed=123, tests = 1:5,}\\
\indent\texttt{alpha = 0.05, control.fdr = TRUE,}\\
\indent\texttt{verbose = FALSE)}\\ 

\noindent\texttt{\# inspect results}\\
\noindent\texttt{unlist(lapply(res, function(df) \{ }\\
\indent\texttt{if("Signif" \%in\% colnames(df)) \{ }\\
\indent\texttt{~~~return(sum(df\$Signif))} \\
\indent\texttt{\} \}))  \# number of significant results}\\ 

Results provided in the heatmaps can be obtained by plotting the p-values using the R package \textbf{pheatmap} \citep{kolde:2019}.

The L1-PCA results were obtained using the R package \textbf{pcaL1}  \citep{Jot:2023}. 
The following code shows how L1-PCA can be run.\\

\noindent\texttt{require(pcaL1)}\\

\noindent\texttt{squashes.L1pca <- l1pca(res\$CATA.table, projDim = 2)}\\ 
\noindent\texttt{plot(squashes.L1pca\$scores, axes = FALSE,}\\ 
\indent\texttt{     xlab = "Component 1", ylab = "Component 2", type = "n")}\\ 
\noindent\texttt{abline(h = 0, v = 0, lty = "dotted", col = "grey")}\\ 
\noindent\texttt{text(squashes.L1pca\$scores, labels = paste0("P", 1:11))}\\ 

After obtaining bootstrap scores, confidence ellipses can be added using the R package \textbf{easyCODA} \citep{Greenacre:18}.

\subsection*{S2. Cochran's Q test results}
Cochran's Q test results are presented in Suppl. Table \ref{SupplTbl_CochranQ}.

\begin{center}
\begin{table*}[h! b]
{\small
\begin{tabular}{lccc}
\textsf{Term} &  \textsf{Cochran Q p-value} & \textsf{BH step-up value}\\
\hline
\textit{Happy} & 0.0000 & 0.0015\\
\textit{Unhappy} & 0.0000 & 0.0029\\
\textit{Uncomfortable} & 0.0000 & 0.0044\\
\textit{At~ease} & 0.0000 & 0.0059\\
\textit{Pleasant~surprise} & 0.0000 & 0.0074\\
\textit{Unpleasant~surprise} & 0.0000 & 0.0088\\
\textit{Disappointment} & 0.0000 & 0.0103\\
\textit{Satisfaction} & 0.0000 & 0.0118\\
\textit{Discontent} & 0.0000 & 0.0132\\
\textit{Trust} & 0.0000 & 0.0279\\
\textit{Disgust} & 0.0000 & 0.0147\\
\textit{Sickly} & 0.0000 & 0.0324\\
\textit{Interested} & 0.0000 & 0.0162\\
\textit{Good} & 0.0000 & 0.0176\\
\textit{Displeasure} & 0.0000 & 0.0191\\
\textit{Comforted} & 0.0000 & 0.0294\\
\textit{Guilty~pleasure} & 0.0000 & 0.0250\\
\textit{Desire} & 0.0000 & 0.0368\\
\textit{Regret} & 0.0000 & 0.0265\\
\textit{Annoyed} & 0.0000 & 0.0206\\
\textit{Pleased} & 0.0000 & 0.0221\\
\textit{Shocked} & 0.0000 & 0.0235\\
\textit{Approval} & 0.0001 & 0.0309\\
\textit{Bored} & 0.0005 & 0.0441\\
\textit{Angry} & 0.0015 & 0.0412\\
\textit{Resentment} & 0.0024 & 0.0338\\
\textit{Warm} & 0.0030 & 0.0471\\
\textit{Reminiscence} & 0.0102 & 0.0426\\
\textit{Sceptical} & 0.0107 & 0.0353\\
\textit{Curious} & 0.0143 & 0.0382\\
\hline
\hline
\textit{Attentive} & 0.0821 & 0.0397\\
\textit{Cautious} & 0.2919 & 0.0500\\
\textit{Worried} & 0.3305 & 0.0456\\
\textit{Confused} & 0.6573 & 0.0485\\
\hline
\end{tabular}
}
\smallskip
\caption{Cochran's Q test was conducted controlling the false discovery rate at 0.05. Sorted p-values are shown alongside the Benjamini-Hochberg (BH) series. The 30 terms above the double line were below the BH critical value of 0.0143; the other four terms did not discriminate the products.}
\label{SupplTbl_CochranQ}
\end{table*}
\end{center}

In total, Cochran's Q test (Suppl. Table \ref{SupplTbl_CochranQ}) detected differences among products on 30 of 34 terms. The permutation tests based on Test 2 (Table 2) detected differences among products on 22 of these terms. The eight additional terms for which Cochran's Q test detected product differences were \textit{Desire}, \textit{Bored}, \textit{Angry}, \textit{Resentment}, \textit{Warm}, \textit{Reminiscence}, \textit{Sceptical}, and \textit{Curious}. 

\subsection*{S3. McNemar's test results}
McNemar's test results are presented in Suppl. Table \ref{SupplTbl_McNemar}. Controlling the FDR at 0.05 yielded a \emph{lower} BH critical value ($0.0118$) than was obtained in Test 5 ($0.0130$). McNemar's test also detected fewer significant product pairs ($442$) than was reported for Test 5 ($487$) in Section 3.1. 

The same pattern held if the FDR is left uncontrolled. 
McNemar's test detected fewer significant paired comparisons ($639$) than the Test 5 ($660$). 
In this case, Test 5, which was based on permutation testing following the ``one citation, one vote" principle, was more sensitive in detecting pairwise differences than the exact version of McNemar's test, which was based on the two-tailed binomial test. 

\begin{center}
\begin{table*}[h! b]
{\small
\begin{tabular}{lccc}
\textsf{Term} &  \textsf{Signif. Pairs (FDR=0.05)} & \textsf{Significant Pairs}\\
\hline
\textit{Displeasure} &  27 & 30 \\
\textit{Good} &  26 & 32 \\
\textit{Happy} &  25 & 33 \\
\textit{Unpleasant~surprise} &  25 & 31 \\
\textit{Disgust} &  25 & 30 \\
\textit{Satisfaction} &  25 & 29 \\
\textit{Unhappy} &  23 & 28 \\
\textit{Disappointment} &  22 & 30 \\
\textit{At~ease} &  22 & 29 \\
\textit{Discontent} &  22 & 27 \\
\textit{Pleasant~surprise} &  21 & 28 \\
\textit{Pleased} &  18 & 26 \\
\textit{Annoyed} &  17 & 26 \\
\textit{Interested} &  16 & 27 \\
\textit{Uncomfortable} &  15 & 25 \\
\textit{Shocked} &  13 & 19 \\
\textit{Comforted} &  12 & 17 \\
\textit{Regret} &  11 & 18 \\
\textit{Trust} &  11 & 18 \\
\textit{Sickly} &  11 & 17 \\
\textit{Desire} &  10 & 17 \\
\textit{Guilty~pleasure} &  9 & 15 \\
\textit{Approval} &  8 & 15 \\
\textit{Bored} &  8 & 11 \\
\textit{Curious} &  5 & 8 \\
\textit{Warm} &  4 & 11 \\
\textit{Reminiscence} &  3 & 9 \\
\textit{Resentment} &  3 & 9 \\
\textit{Sceptical} &  3 & 5 \\
\textit{Angry} &  2 & 8 \\
\textit{Attentive} &  0 & 5 \\
\textit{Cautious} &  0 & 4 \\
\textit{Worried} &  0 & 2 \\
\textit{Confused} &  0 & 0 \\
\hline
\end{tabular}
}
\smallskip
\caption{McNemar's test of each term on 55 paired comparisons (1870 tests) yielded 442 significant test results when the false discovery rate was controlled at 0.05 (``Signif. Pairs (FDR=0.05)"); otherwise, 639 test results are significant (``Signif. Pairs").}
\label{SupplTbl_McNemar}
\end{table*}
\end{center}

\subsection*{S4. Principal component analysis and correspondence analysis}
In this section, we will compare the L1-PCA used in this paper with the conventional PCA which uses the L2 norm, referred to as L2-PCA. We will also discuss different ways of conducting correspondence analysis of a CATA frequency table. Results based on the CATA table of blackcurrant squashes will illustrate the differences between methods.

\begin{figure*}[ht]
\begin{center}
\includegraphics[width=11.8cm]{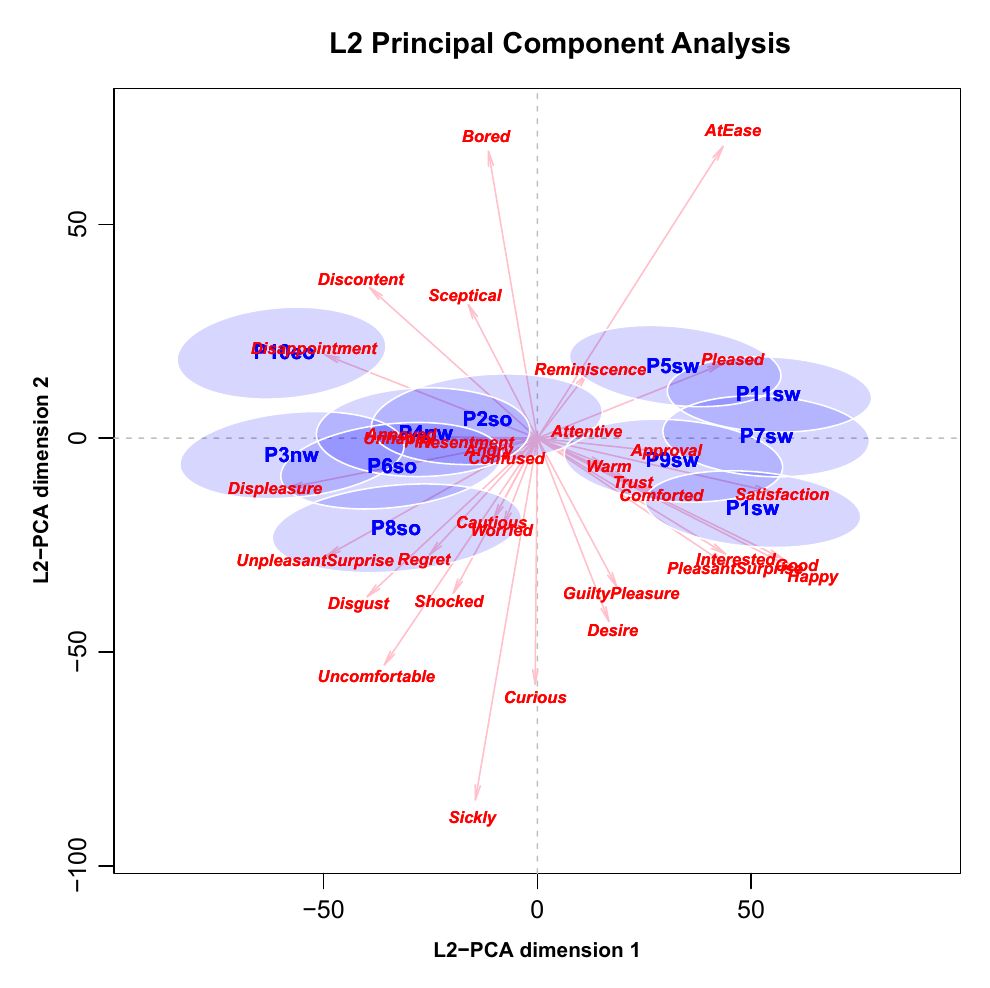}
\caption{L2-PCA of the CATA table, with confidence ellipses for the product positions. Percentages of explained variance on these two dimensions are 82.3\% and 6.6\% respectively.}
\label{SupplFig_L2PCA}
\end{center}
\end{figure*}

In the L1-PCA, used in the ``one citation, one vote" approach, the deviation of each cell $e_{pt}$ of the CATA table {\bf E} is measured with respect to the median \textsf{M}($t$) by the absolute difference $\vert e_{pt} - \textsf{M}(t)\vert$ (see Eq.~(1) of the paper). 
In L2-PCA, the deviations are measured by squared differences with respect to the term mean $\bar{e}_t$: $(e_{pt} - \bar{e}_t)^2$, not with respect to the median.
Writing this squared difference as $\vert e_{pt} - \bar{e}_t\vert \cdot \vert e_{pt} - \bar{e}_t\vert$, shows that the absolute difference with respect to the term mean, $\vert e_{pt} - \bar{e}_t\vert$, can be considered weighted by exactly the same amount. 
This weighting increases the contribution of large deviations from the mean compared to small deviations in the measure of total dispersion in the data table, whereas our L1-norm approach leaves such deviations (from the median) unweighted.
Nevertheless, the two-dimensional L2-PCA solution of the blackcurrant squashes table, shown in Fig. \ref{SupplFig_L2PCA}, is quite similar to that of the L1-PCA result of Fig. 7.
Apart from slightly different emphases on some terms, the most noticeable difference is the increased elongation of the confidence ellipses along the first dimension, which accounts for a high percentage (82.3\%) of the variance in the table of citation percentages.
This high percentage compared to the lower percentage of 58.6\% of the first dimension in the L1-PCA is a result of the squaring of the deviations, a convex function that exaggerates the higher dispersions. 
Notice too that in the L2-PCA, as in the L1-norm version, the term citation values are not standardized, since the natural variances of the terms are required to be preserved, not equalized.

Correspondence analysis (CA) is a popular dimension-reduction method for tables of counts, and inherent in CA is the chi-square normalization.
This normalization divides each squared difference between the observed count $e_{pt}$ and respective term mean count $\bar{e}_t$ by the inverse of the mean count: $(e_{pt} - \bar{e}_t)^2/\bar{e}_t$, which is the typical chi-square form of (observed-expected)$^2$/expected.
This quantity can be written as $\vert e_{pt} - \bar{e}_t\vert \cdot \vert e_{pt} - \bar{e}_t\vert / \bar{e}_t$.
So CA, being an L2-norm method, not only ``boosts" each absolute difference with respect to the mean by the same absolute difference, but then additionally weights by the inverse of the mean, which has an effect of downweighting the higher counts.
Looked at it this way, CA can be thought of as having weights on each deviation $\vert e_{pt} - \bar{e}_t\vert $ equal to $\vert e_{pt} - \bar{e}_t\vert / \bar{e}_t$, where the numerator and denominator in this weight ratio can sometimes partially cancel out, since terms with larger means will tend to have larger deviations from the mean.
Nevertheless, it is clear that this weighting (or any other weighting of the terms) introduces varying numbers of ``votes" for each citation, which disagrees with our ``one citation, one vote" principle.

However, there is a problem applying regular CA to a CATA table, which is not a contingency table. 
Using CA in this case has to be done with careful consideration of a major property inherent in the method, namely the expressing of the row (or column) counts relative to their respective totals.
For example, the set of row (product) counts is converted to a set of relative counts, called profiles or compositions, and it is these relative counts that are being visualized by CA.
These relative counts depend on the total number of citations for each product, which vary.
\cite{mahieu2021multiple} have quite correctly drawn attention to this point and proposed an alternative analysis, multiple response CA (mrCA), which does not change the counts into relative counts, but still conserves the chi-square style of standardization.
It turns out that their mrCA can be equivalently computed using either a variant of CA called ``subset CA" \citep{GreenacrePardo:06} or another version called ``CA of raw data" \citep{Greenacre:10c}.

Subset CA analyses a selected set of rows and/or columns in a table but maintains the margins of the original table.
To keep the original CATA table values and not relativize them with respect to their row totals, a dummy column is added to the CATA table to force the table to have equal row totals. 
For example, the dummy column could be, for each row $p$, $\sum_t (100-e_{pt})$, i.e. the sum of the percentages of non-citations in the CATA table. 
Then the row sums of the new $P\times (T+1)$ table are a constant $100T$.
The subset CA is performed indicating that the subset of columns of interest consists of the first $T$ columns, omitting the dummy column introduced to force equal row margins.

The CA of raw data, which can be called rawCA, originated independently in the context of ecological abundance data, where the actual abundance levels were of interest, not their relative abundances that regular CA analyses by default. 
Thus, in this analysis the inherent proportioning out of the counts with respect to the totals is suppressed, and the raw data are analysed, not their relative values.

It turns out that the solutions of both mrCA and rawCA are equivalent to the subset CA solution, which was also developed in a completely different context, namely for analysing survey data where specific response options were required to be visualized.

The subset CA analysis can be simply achieved with the R package {\bf ca} \citep{NenadicGreenacre:07}, using the option \texttt{subsetcol}, as follows:\\

\noindent\texttt{\# Subset CA to perform Mahieu et al.'s CATA analysis }\\
\noindent\texttt{\# Suppose CATA table of percentages is in E, with T columns}\\
\noindent\texttt{require(ca)}\\
\noindent\texttt{E2 <- cbind(E, 100*T - rowSums(E))}\\
\noindent\texttt{E.subCA <- ca(E2, subsetcol=1:T)} \\
\noindent\texttt{plot(E.subCA)}\\

\begin{figure*}[ht]
\begin{center}
\includegraphics[width=11.8cm]{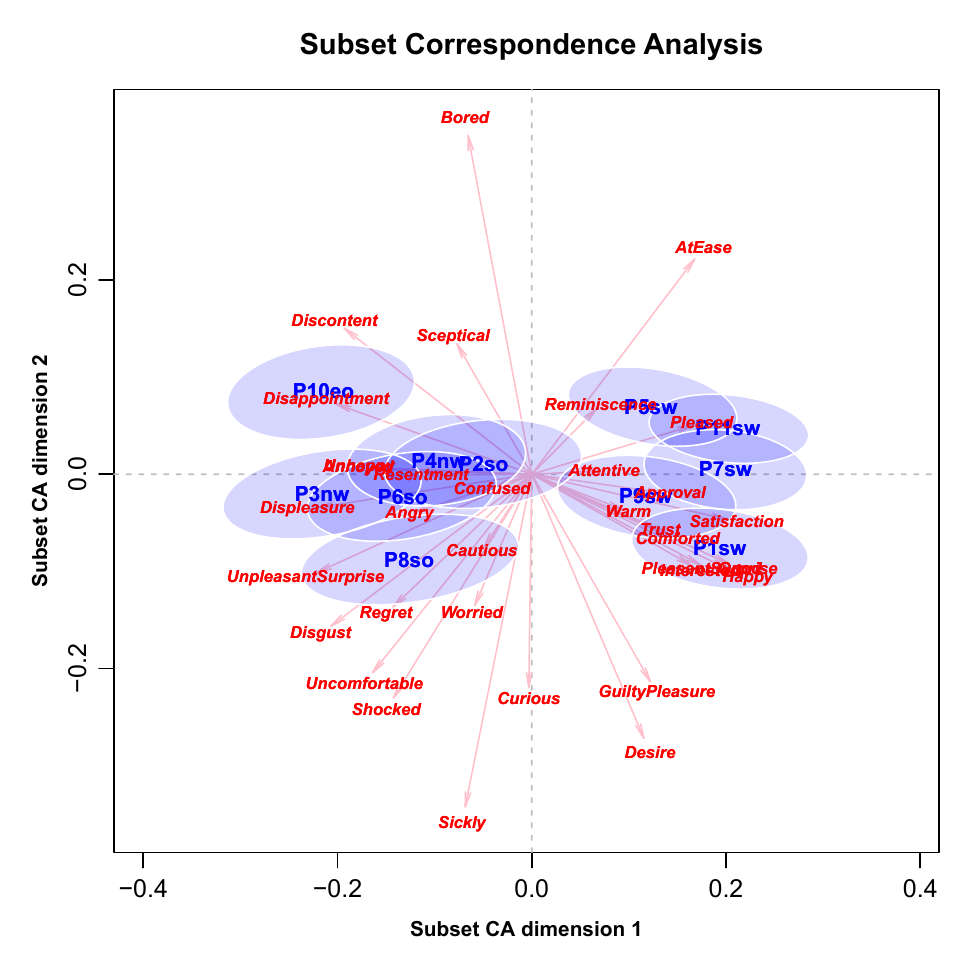}
\caption{Subset CA of the CATA table, with confidence ellipses for the product positions. Percentages of explained variance on these two dimensions are 78.3\% and 8.2\% resoectively.}
\label{SupplFig_subCA}
\end{center}
\end{figure*}

The result of the above code, applied to the CATA table, and enhanced with the confidence ellipses, is given in Fig. \ref{SupplFig_subCA}.
Again, there are only minor differences compared to the L1-PCA biplot (Fig. 7) and the L2-PCA biplot (Fig. \ref{SupplFig_L2PCA}).
However, these analyses might not be similar in other applications.

\subsection*{S5. Three-dimensional view of the L1-PCA solution}
Suppl. Video V1 shows an animation of the three-dimensional L1-PCA solution, rotating by 360 degrees around the vertical second axis. 
The ellipsoids of the products that are standard with added sugar are coloured red, to show the contrast with the other products on the first dimension. 
The video pauses at the initial dimension 1 by dimension 2 view and then again at the dimension 3 by dimension 2 view.
In the first pausing, the non-overlapping confidence ellipsoids show the products are well separated, but in the second pausing, all confidence ellipsoids overlap on the third dimension, showing that the products are not well distinguished in the third component. We focus our interpretation on the two-component L1-PCA, which differs slightly from the first plane of the three-component L1-PCA solution due to the non-nestedness of L1-PCA.
\end{document}